# THE CATALYTIC ACTIVITY OF GOLD/CADMIUM SULFIDE (Au/CdS) NANOCRYSTALS

Ebin Bastola

A Thesis

Submitted to the Graduate College of Bowling Green
State University in partial fulfillment of
the requirements for the degree of

MASTER OF SCIENCE

August 2014

Committee:

Mikhail Zamkov, Advisor

Alexy Zayak

Farida Selim

# ABSTRACT

Mikhail Zamkov, Advisor

Semiconductor nanocrystals (NCs) are the building blocks for the future opto-electronic devices, and other energy-harvesting applications. Although the energy demand has been increasing, energy sources are diminishing. Hydrogen, green energy source, is used in space shuttles and other transportation means. The hydrogen production using semiconductors such as $TiO_2$ suffers from several limitations, and hence, alterative techniques for hydrogen production are desired. Recently, photocatalysis using different nanostructures has been studied.[1-6] In this project, we have studied the catalytic activity of gold/cadmium sulphide (Au/CdS) NCs to reduce protons into hydrogen qualitatively. The Au/CdS NCs absorb a large amount of visible photons due to plasmon resonance and create electron-hole pairs. The excited electrons migrate to the catalysts, and the protons are reduced while the ligands on the surface of NCs scavenge holes. Thus, photocatalysis occurs due to the energy transfer from metal to semiconductor materials. The hydrogen production using Au/CdS metal semiconductor NCs is more efficient than that of using CdS semiconductor NCs.

I

would like

to

dedicate

this work to

my father

KHAGENDRA PRASAD BASTOLA

and

my mother

TULASA BASTOLA

who

always encourage

me

for higher

study and research.

# ACKNOWLEDGMENTS

I would like to thank my supervisor, Dr. Mikhail Zamkov, for the support and guidance he has given me during the research work in his lab. I am delighted to have been a member of his research group during my study at Bowling Green State University; it has opened a new way to my life for further study and research. Also, I would like to thank Dr. Elena Khon for her help when I joined the research lab and during the writing of this thesis.

Further, I would like to thank Dr. Alexy T. Zayak and Dr. Farida A. Selim, my committee members, for spending their time to review my thesis.

I cannot remain without thanking Dr. Lewis P. Fulcher, Graduate Coordinator of the Department of Physics, for his support, care, and guidance during my Master's Degree study at Bowling Green.

Additionally, I would like to thank Angela Garner, English Teacher in the ESOL program, for helping me to minimize the language errors.

Finally, I am extremely thankful to my parents and wife for their support, love, and care which always inspired me to advance in my study and research.

# TABLE OF CONTENTS

# LIST OF FIGURES

# CHAPTER I: INTRODUCTION

Semiconductor nanocrystals (NCs) are crystalline particles with sizes ranging from 1 nm to 10 nm composed of a few hundreds to thousands of atoms. The size of the semiconductor NCs is important to determine their optical properties. For example, when the size of the CdSe NCs grows from a diameter of 2 nm to 6 nm, the emission color changes from blue to red.[7] This emission of different colors with different sizes is due to the exciton confinement. It can be explained by an example of a particle in a box. When the size of the box increases, the energy levels are close to each other with a slight difference in confinement, and the smaller size of the box creates a large separation of energy levels leading to a stronger confinement. This color emission with increase in NCs sizes can be understood by a concept of combining orbitals.

These NCs are also considered to be intermediate particles between atoms or molecules and bulk materials. The energy levels of a simple molecule can be calculated by using the Linear Combination of Atomic Orbitals (LCAO) method.[8] In this approach, the outer orbitals of atoms form molecular orbitals with different energy levels. In general, this LCAO approach produces two molecular orbitals: one orbital, bonding, has less energy level than the atomic orbitals, and another, antibonding, has higher energy level. When this approach is extended to a large number of atoms, it forms a band. The binding orbitals form a valence band, and antibonding orbitals form a conduction band.[9] The gap between the bands (ie. bandgap or energy gap) decreases with the large number of participating atoms. Thus, the larger the size of the NCs, the greater is the shift in the fluorescence spectra towards the red light as in CdSe NCs.

Moreover, the combination of metals with semiconductor NCs forms metal semiconductor NCs. Colloidal metal semiconductor NCs have similar optical properties to semiconductor NCs. However, the absorption spectra in metal semiconductor NCs is not due to transitions between discrete energy states. Rather, when the frequency of incident light radiation matches the frequency of the surface electrons on the metal semiconductor NCs surface, then it excites surface electrons. The excitation of surface electrons in metal semiconductor NCs with the incident of suitable light frequency is called surface plasmon resonance. Hence, there is a collective excitation of the electron gas in the metal NCs. The peak on the absorption spectra of metal NCs is the resonance frequency of surface plasmons and the surface plasmon resonance in nanometer scale is called localized surface plasmon resonance (LSPR).[10] Thus, because of this LSPR on the metal semiconductor NCs, the electric field near the surface of metal semiconductor NCs is significantly improved. The plasmon excitation provides an opportunity for enhancing the optical absorption of metal semiconductor nanocomposites through energy transfer. The nonepitaxial metal semiconductor nanocomposites can be used to develop hybrid nanoscale systems that encourage exciton-plasmon interaction without unwanted side reactions and defect-localized energy dissipations.[11]

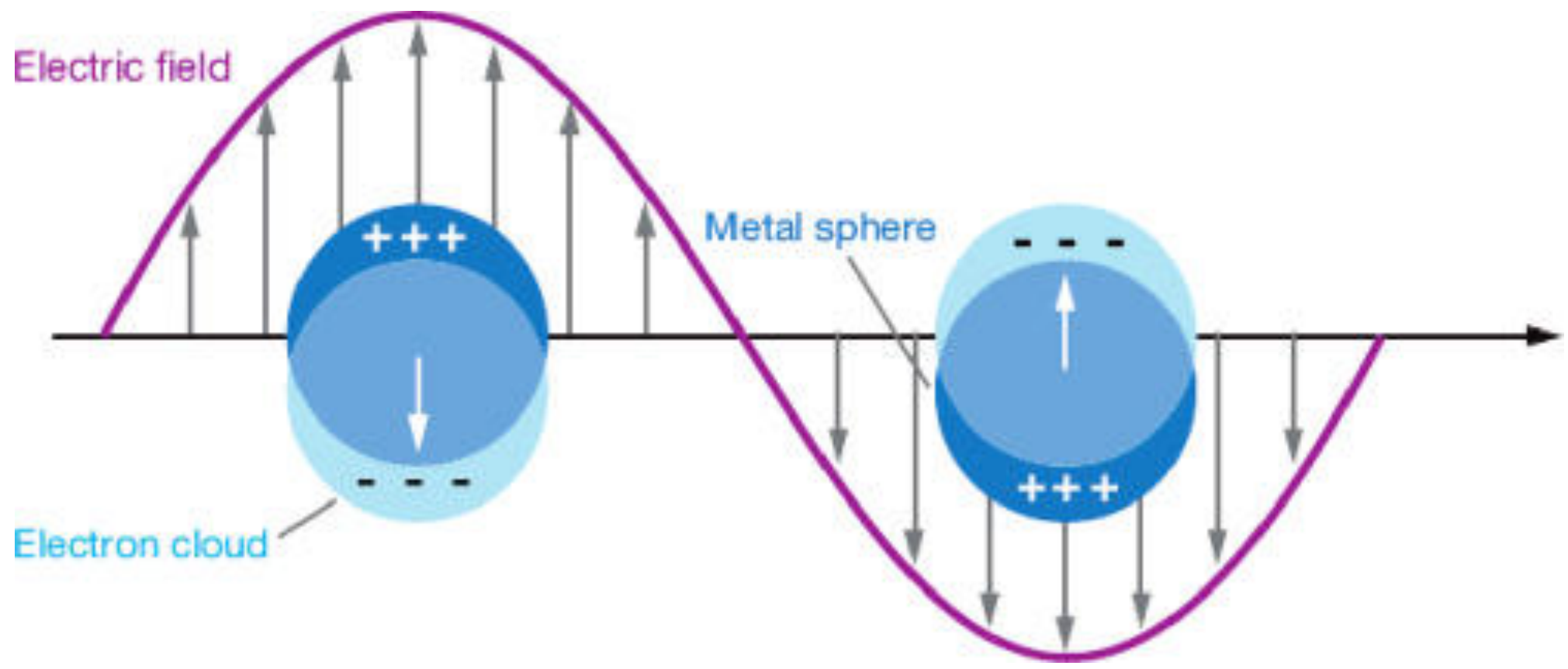

Figure 1.1 Surface Plasmon enhancement in Metal Semiconductor NCs [12]

The combination of two different types of semiconductor NCs yields two different NCs structures: Type I and Type II. In Type I, the NCs with a smaller bandgap are surrounded by another semiconductor material of a larger bandgap as in CdSe/CdS and CdSe/ZnS Core/Shell structures. Similarly, if the large bandgap semiconductor NCs are surrounded by a smaller bandgap materials, it is called reverse Type I. For example, in CdS/CdSe and CdS/HgS cores/shells structures, the larger bandgap material, CdS (2.42 eV), is surrounded by smaller bandgap materials, CdSe (1.74eV) and HgS (2.1 eV), respectively (See Figure 1.2). Further, in Type II structures, the valence and conduction band edges of the core material NCs are both higher or lower than the band edges of the shell material NCs. For example, ZnTe (2.26 eV) is surrounded by CdSe (1.74eV) (See Figure 1.3) forms Type II semiconductor NCs structure.

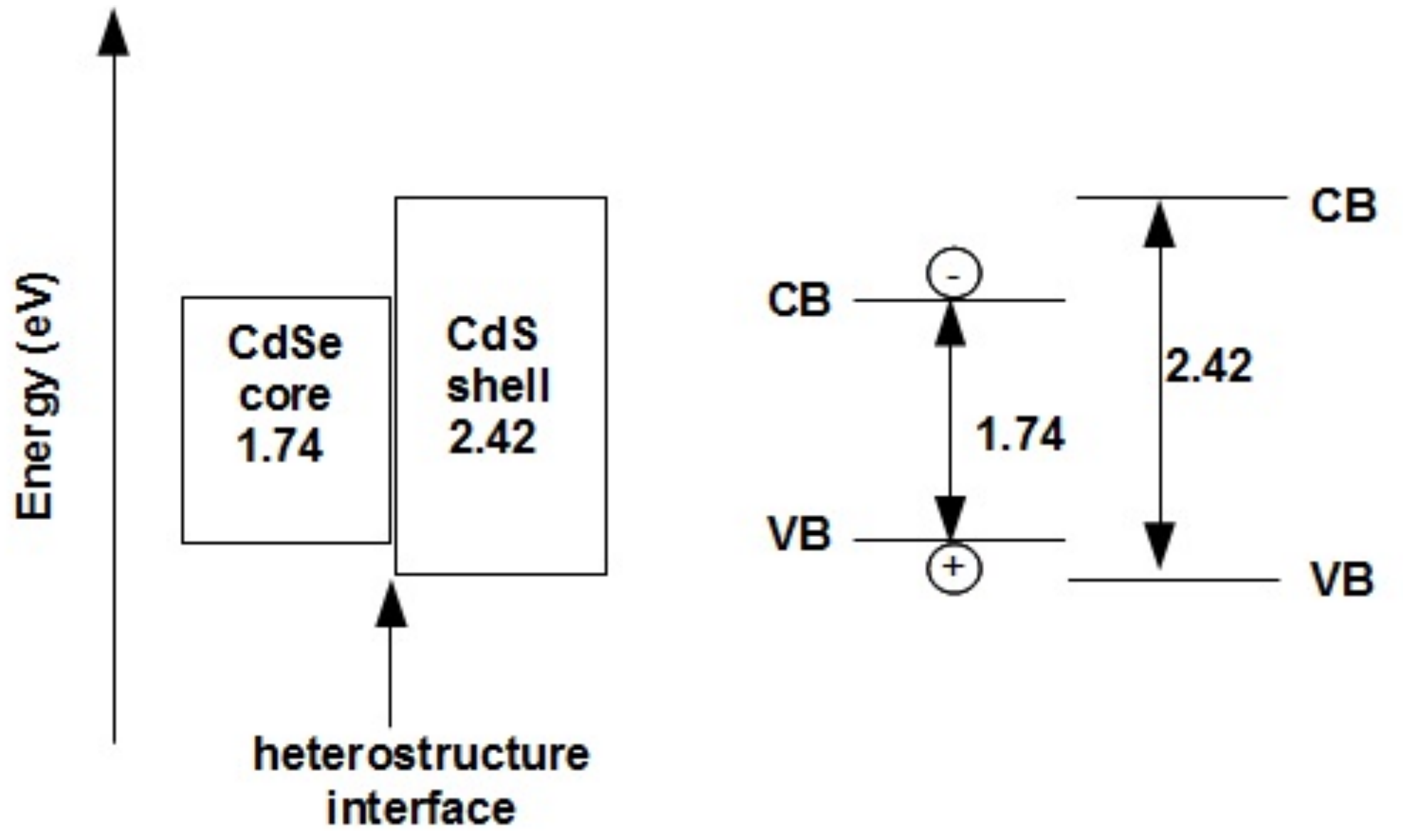

Figure 1.2 Energy Bandgap Diagram of CdSe/CdS Core/Shell NCs Type I Structure [13]

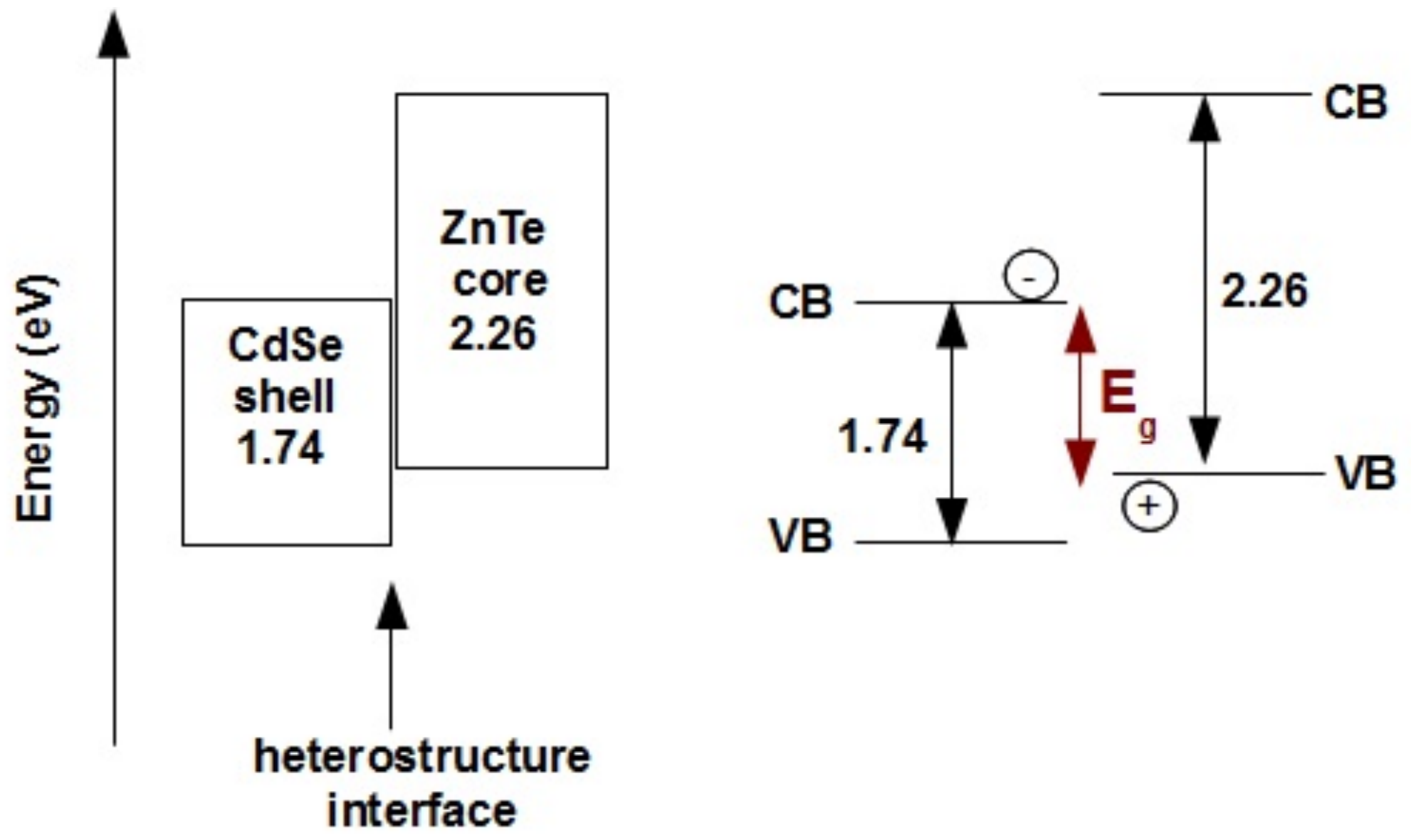

Figure 1.3 Energy Bandgap Diagram of CdSe/ZnTe Core/Shell NCs Type II Structure [14]

Due to their unique properties and various synthesis methods for controlling their shape and size, NCs are promising materials for the development of optical and electronic devices. The control over the size and shape of the NCs opens a door to a new research field. These NCs are widely used in different applications such as light emitting devices, photodetectors, biomedical imaging, medical diagonstics, solar cells, and photocatalysis.

The optical and electrical properties make NCs unique materials for light emitting devices and photodetectors. The liquid crystal display (LCD) is almost replaced by light emitting diodes (LED) which produce red, green and blue pixels after color filter. The combination of optical properties of NCs with the electrical properties of conductive polymers make promising materials for photonic applications that significantly reduces the cost than the existing semiconductor devices. Likewise, semiconductor NCs, such as CdSe, can be used to make optical temperature sensors since the photoluminescence intensity decreases linearly with increasing temperature, and the emission spectra shifts towards red with rise in temperature.[15]

Because of their optical properties, resistance for photobleaching, and tunable wavelength, semiconductor NCs have a great potential for biomedical imaging and medical diagnostics of different diseases including cancer. The broad absorption band of NCs dominates the narrow absorption band of traditional organic dyes and opens the door for potential advances in biomedical imaging. [16, 17] The use of quantum dots (QDs) for diagnosing diseases is recently the most advanced in the field of cancer treatment. Cancer is a disease killing million people in a year. The existing cancer-detecting techniques including tissue biopsies, medical imaging, and bioanalytic assay of urine and blood are not sensitive enough to detect the early tumor formation. These techniques can only detect the solid tumors of 1 cm in diameter, at that point, the tumor consists of millions of cells, and might have been metastasized.[18] Thus, to overcome this diagnosis difficulty, QDs with their unique optical properties can be used as high contrast resolution agents for biomedical imaging of the smallest tumors and to screen cancer markers in urine as well as blood and tissue biopsies.[19] These experiments that use QDs to identify the tumors in animals show promising results in possible applications for human beings.[20]

Today, the great challenge to researchers is to address the problem of increasing energy demand and diminishing energy supply. Scientists have been trying to develop alternative sources of energy that are renewable, environment friendly and carbon free. Solar cells are one of the alternatives of existing fossil fuels to harvest solar energy. Scientists have been investigating different materials, for example semiconductor NCs, for the solar cells that can absorb solar radiation and produce energy at low cost. Recently, researchers reported that single-junction power-conversion efficiencies of 6% by using colloidal quantum dots in depleted-heterostructure solar cells.[21] These nanocrystal solar cells have better thermal and chemical stability against heat and oxygenated environments.[22] Since the nanocrystal solar cells are based on colloidal NCs, it is less expensive to fabricate a device than to use existing semiconductors in solar cells. Another approach to get clean energy is direct conversion of solar energy into chemical energy via photocatalysis.

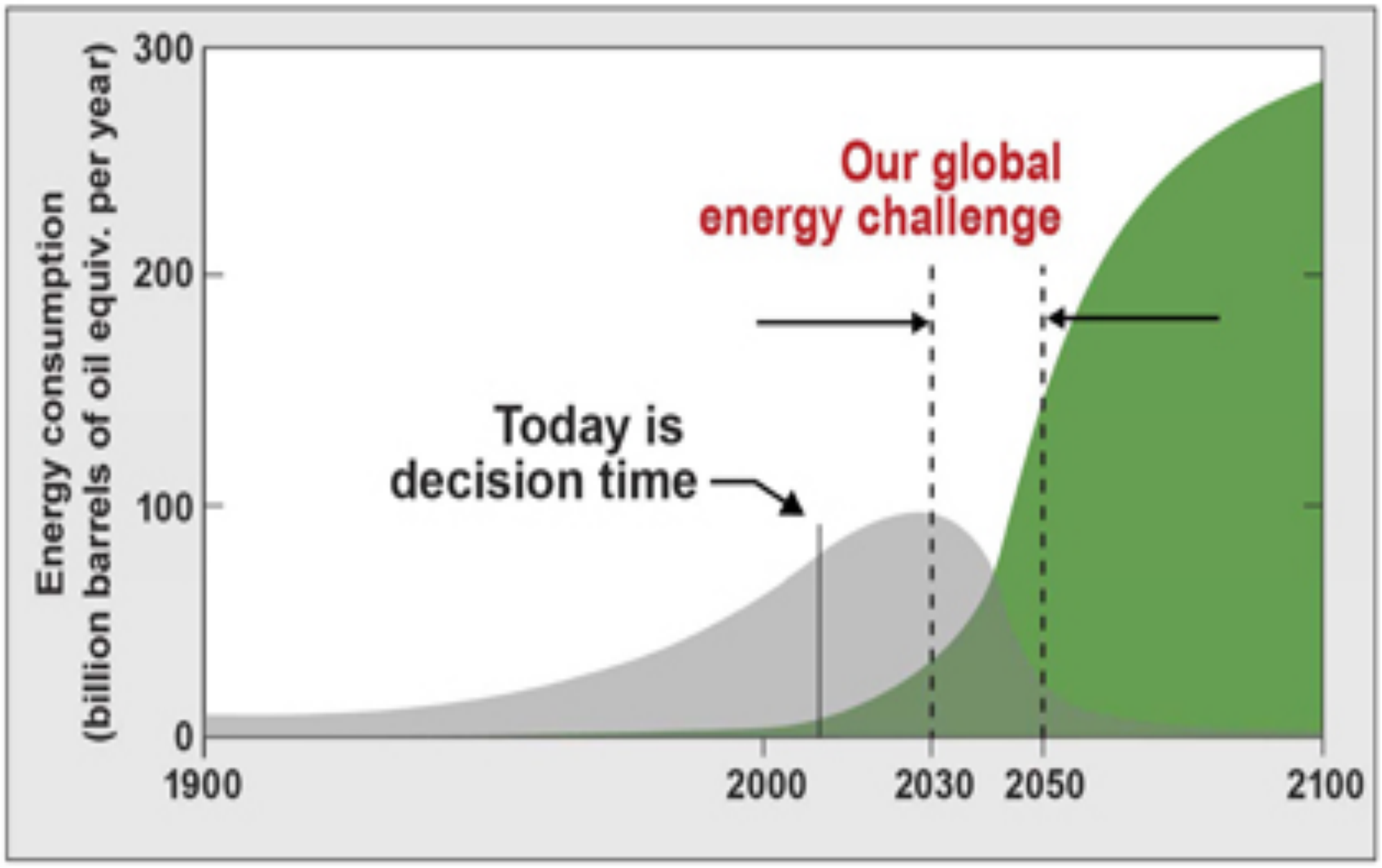

Figure 1.4 Energy Supply and Demand Current and Future Projection [23] (The grey part represents the diminishing energy supply and, the green part indicates accelarating energy demand.)

Photocatlysis is the process of conversion of light energy into the chemical energy in the presence of a suitable chemical reagent called photocatalyst. The catalysts itself does not change during the chemical reaction, but it accelerates the rate of a reaction. There are two types of photocatalytic reactions: Homogeneous Photocatalysis and Heterogeneous Photocatalysis. Homogeneous photocatalytic reactions occur when the reactants are in same phase and heterogeneous photocatalytic reactions occur when the reactants are in different phases.

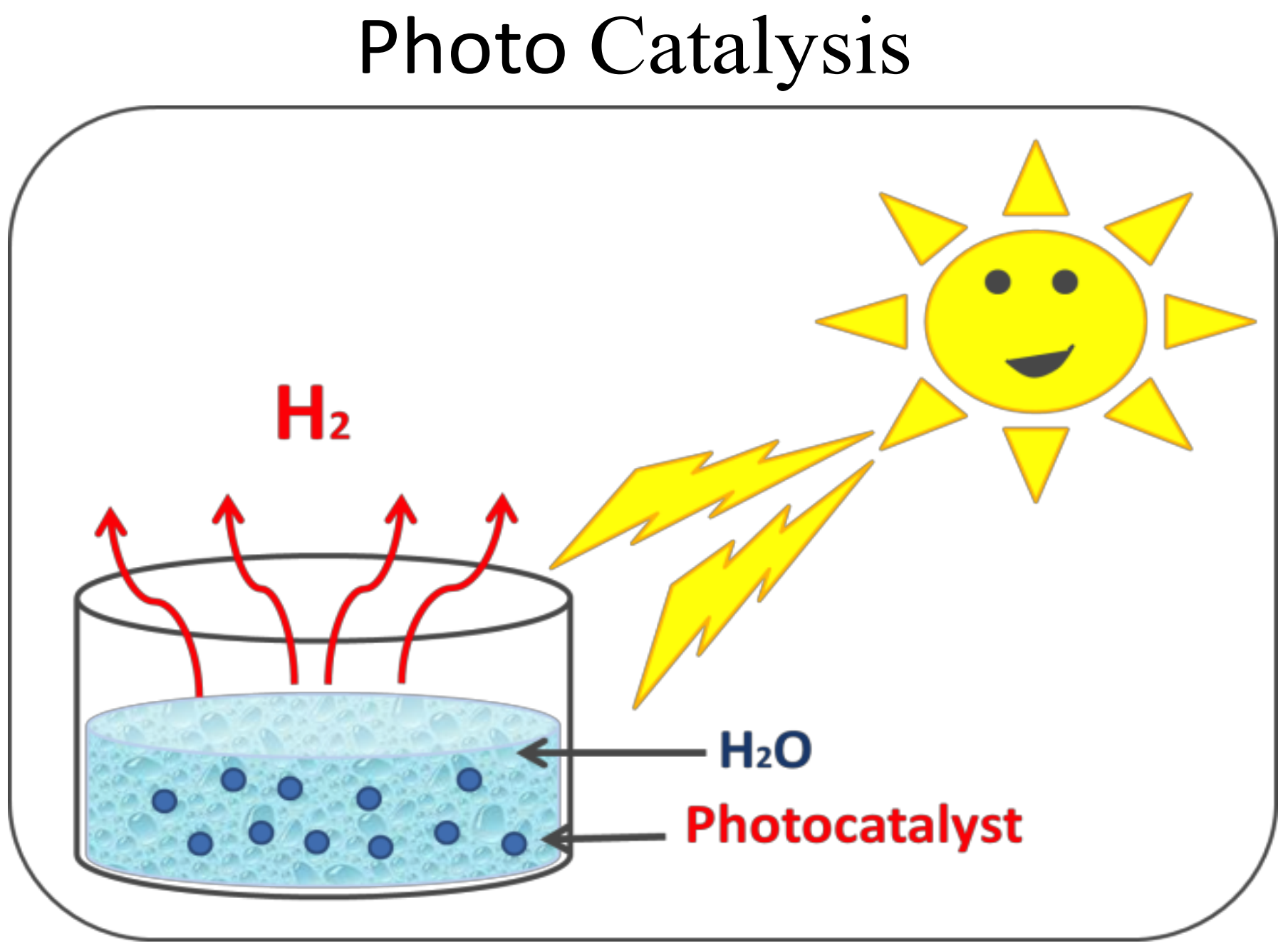

Figure 1.5 Illustration of Photocatalysis in Presence of Sunlight

Researchers have paid much attention to photocatalysis to produce clean energy and minimize the environmental problems. To address the energy crisis, the semiconductor photocatalysis is used to produce hydrogen, a clean energy source. Furthermore, photocatalysis can be employed to reduce air and water pollution. Daneshvar et al. reported that ZnO can be used to degrade the insecticide diazinon in presence of UV-light. In their study, the authors found that prepared NCs offered electrical energy efficiency and better quantum yield than the commercially available ZnO.[24] It can also be useful to reduce the spread of infections and diseases as well as to clean houses, living environments and other facilities.

The narrow bandgap between the valence and the conduction band makes semiconductor photocatalysts useful in semiconductor photocatalysis. Fujishima and Honda first discovered the splitting of water into hydrogen and oxygen using a semiconductor ($TiO_2$) catalyst under UV irradiation.[25] In recent years, other semiconductors including semiconductor NCs and metal semiconductor nanostructures (CdSe, Au/CdS) have been used to breakdown water into hydrogen and oxygen.

Light plays a vital role to initiate the photocatalytic reactions on the surface of semiconductors. As light incidents on the surface of a semiconductor catalyst, it generates the electron-hole pairs. To produce these electron-hole pairs, the photon energy should be greater than the bandgap energy required to move an electron from the conduction band to the valence band. When the excited electrons move to the surface of the semiconductor, they can take part in oxidation and reduction reactions.

$$TiO_2 \longrightarrow e_{cb}^{-} + h_{vb}^{+} \quad \text{(In Presence of Light)}$$

The photocatalytic mechanism that uses a semiconductor catalyst can be understood from the schematic diagram shown below.

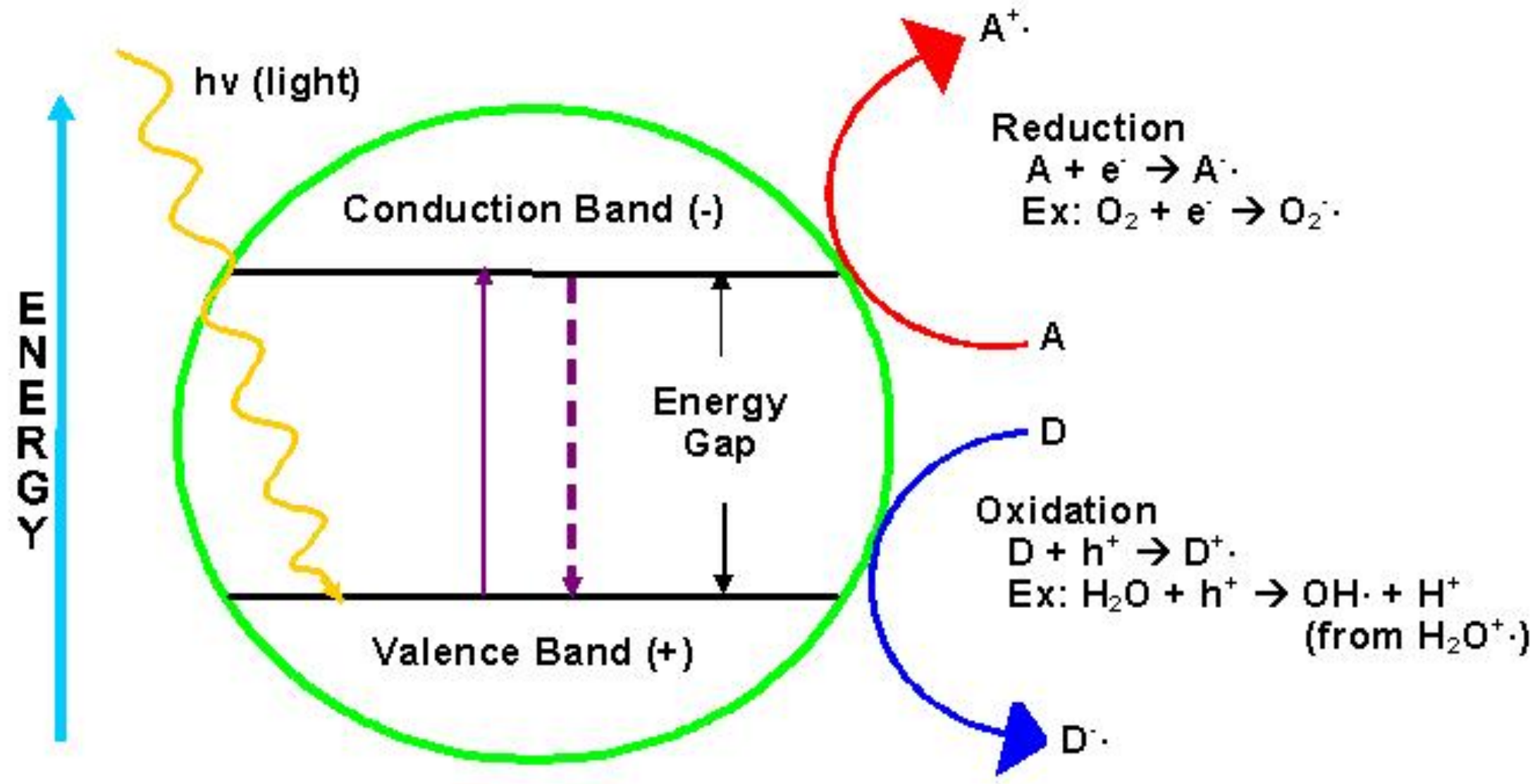

Figure 1.6 Schematic Representation of Semiconductor Photocatalysis

The reactions for the splitting of water into hydrogen and oxygen can be summarized below.

Reduction: $2\ H^+(aq) + 2e^- \rightarrow H_2(g)$

Oxidation: $2\ H_2O(l) \rightarrow O_2(g) + 4\ H^+(aq) + 4e^-$

Overall reaction: $2\ H_2O\ (l) \rightarrow 2\ H_2(g) + O_2(g)$

Here, the hydrogen produced is significant since it is used as a clean energy source in space shuttles and other transportation means. The reactions occur when the photons have an energy greater than 1.23 ev. These reactions proceed infinitesimally slow in the presence of light only, and hence we need to have a suitable photocatalyst that accelerates the reaction rate in forward direction.

Since $TiO_2$ (Eg = 3.2 ev) is inert and less corrosive, it is the most commonly used semiconductor photocatalyst. However, ultraviolet light is required for photocatalysis to take place. The band gap of $TiO_2$ can be narrowed to absorb visible photons. Chen et al. introduced a conceptually different approach in which disorder-engineered $TiO_2$ NCs possess substantial solar-driven photocatalysis for the production of hydrogen due to mid-gap electronic states which reduces the band gap of $TiO_2$.[26] Waterhouse et al. reported that tuning the photonic band gap to electronic band gap of Au/ $TiO_2$ catalysts enhances the photocatalytic splitting of water under direct sunlight.[27] Another study of $ZnFe_2O_4$ photocatalytic activity demonstrated that reduction of water into hydrogen under visible light (≥ 420nm ) is much higher than the $TiO_2$-xNx photocatalyst due to the small band gap of the chemically prepared ferrite photocatalyst.[28]

In recent years, the researchers have been using other semiconductors as photocatalysts that can absorb light in visible or infrared regions and produce large amounts of hydrogen after photocatalysis. The crystalline $Zn_2SnO_4$, synthesized by solvothermal routes, has a better photocatalytic property due to its small crystal size and large surface area.[29] The simplicity and potential low cost operation of hydrogen production make the splitting of water of particular interest to the research community. However, using semiconductor catalysts for this process has its own limitations. For example, these catalysts have low efficiency of around 0.1%. The authors stated that using Cobalt (II) oxide (CoO) NCs can improve the efficiency to around 5% for solar-to-hydrogen production.[30]

Recently, scientists have been investigating the photocatalytic activity of different nanostructures. The unique optical property of NCs make them promising material to absorb photons in visible and infrared regions, and hence carry photonic degradation of water into its components. The charge separation in these semiconductor NCs make them useful for the study of the photocatalytic activity. A rod-in-a-dot semiconductor domain (ZnSe/CdS/Pt) in a colloidal heterogeneous system can induce the catalytic reactions as the light-driven hydrogen production by localization of both charges in non-adjacent parts of the nanostructures.[31] In addition, other alternative processing strategies enhance the catalytic activity of nanocrystal architectures. Khon et al. reported that the CdSe/CdS heterostructures can be used to enhance the catalytic activity by 3/4 times after etching treatment due to the separation of charges in the heterostructured NCs.[32] The most significant advantage of using the semiconductor-metal heterostructured NCs is they can be reused for the division of water into hydrogen simply by recharging these nanoparticles with ligands.[4]

The researchers have been trying to develop a renewable, carbon free energy source for everyday use by directly converting solar energy into chemical fuels. Scientists are also studying the catalytic behaviors of semiconductor NCs such as CdSe, since they absorb light in the visible region. The quantitative analysis of quantum-size-controlled photocatalytic $H_2$ production at the semiconductor-solution interface depends on the size of CdSe quantum dots.[33] Recently, a homogeneous system of CdSe capped with dihydrolipoic acid in water to reduce protons into hydrogen is reported in the presence of a nickel catalyst. Here, the CdSe NCs absorb light to create electron-hole pairs, and produce hydrogen. Han et al. reported that under appropriate conditions, NCs achieve quantum yields in water more than 36% under an illunimination of 520 nm light. [1]

This project addresses the photocatalytic activity of Au/CdS metal semiconductor NCs. Moreover, we compare the catalytic behaviors of Au/Cds metal semiconductor NCs with the CdS semiconductor NCs under identical conditions. The advantage of using the metal semiconductor NCs is to absorb large amount of visible photons with less amount of metal semiconductor NCs and generate the large number of electron-hole pairs. The electrons in the conduction band migrate to catalysts available on the surface of Au/Cds NCs and reduce protons into hydrogen whereas the holes move towards the surface ligands and carry out oxidation reactions. Hence, the electrons and holes generated accomplish photocatalytic degradation of water into hydrogen and oxygen. This study shows that photocatalysis using Au/CdS is a reliable and promising method for the production of hydrogen.

# CHAPTER II: EXPERIMENTAL

## *Materials*

1- Octadecene (ODE, 90%; Aldrich), oleylamine (OLAM, 70%; Aldrich), oleic acid (OA, 90%; Aldrich), Acetone (ACS, Amresco), toluene (anhydrous, 99.8%; Aldrich), octane (anhydrous, 99%; Aldrich), silver nitrate (99%, Sigma- Aldrich), gold(III) chloride (99%, Acros Organics), silver nitrate (99%, Sigma-Aldrich), cadmium nitrate tetrahydrate (99.999%, Aldrich), 3-mercaptopropionic acid (3-MPA, 99%; Alfa Aesar), sulfur (S, 99.999%; Acros), tributylphosphine (TBP) (97%, Aldrich), methanol (anhydrous, 99.8%, Aldrich), ethanol (anhydrous, 95%, Aldrich), and toluene (anhydrous, 99.8%, Aldrich), chloroform (99 %, Sigma Aldrich), hexane (anhydrous, 95%, Aldrich), were used as purchased. Cadmium oxide (CdO, 99.99%, Aldrich), 11-mercaptoundecanoic acid (MUA, 95%, Aldrich), platinum (II) acetylacetonate (97%, Aldrich), 1, 2 hexadecane diol (TCI), Asrboic Acid (Sigma), nickel (II) nitrate hexahydrate (99.999 %, Aldrich), potassium hydroxide (KOH; Acros Organics). The reactions were done under argon atmosphere using the standard air free Schlenk technique unless otherwise stated, and the centrifuge used for precipitation rotates at 7200 rpm.

## *Synthesis of Au Cores*

To synthesize the Oleyamine-capped Au cores, the procedure[34] developed in our group was followed. Initially, 0.011 g $AuCl_3$ was taken in one-neck flask and it was dissolved in 3 mL degassed oleyamine (OLAM) by sonication for 10 to 15 minutes. The reaction mixture color changed to orange due to the formation of Au-oleates. Then, the reaction temperature was increased and kept at 100 $^0$C while stirring under argon for 30 minutes. The reaction mixture color changed from orange to purple. It was then cooled to room temperature. The reaction mixture was transformed into two centrifuge tubes, and excess ethanol was added. After that, the tubes were centrifuged for 5 minutes. The supernatant was poured off and the precipitate was dissolved in 6 mL toluene. This cleaning with toluene/ethanol was repeated. Finally, Au cores were collected in 4 mL of toluene.

*Synthesis of Au/Ag Core/Shell NCs*

The procedure reported by Shore et al. was followed to synthesize the oleyamine-capped Au/Ag core/shell NCs.[35] The Au-cores prepared in the previous step was taken in a one-neck flask with 5 mL degassed OLAM. The flask was heated at 120 $^0$C while stirring under argon so that all the toluene was evaporated. The $1.05*10^{-2}$ M $AgNO_3$ solution was prepared separately in deionised water. The $AgNO_3$ solution was sonicated to dissolve it completely. Then, 0.2 mL $AgNO_3$ was added dropwise to the Au cores flask. It was left for 10 minutes at 120 $^0$C to grow the Ag shell over the Au cores. Here, the excess $Ag^+$ ions would produce isolated Ag nanoparticles, so the $AgNO_3$ solution was added very solwly dropwise to avoid the formation of isolated Ag nanoparticles.

The thickness of the Ag shell could be checked by the localised surface plasmon resoance (LSPR) peak absorption. On using all the amount of Au cores made from the previous step, the addition of 0.5 mL $AgNO_3$ would produce a blue-shift of 15 to 20 nm, and 0.9 mL $AgNO_3$ would result a 40 nm blue-shift. Thus, the additional $AgNO_3$ solution would give the desired shift of the LSPR absorption peak. After obtaining the desired shift (i.e., Ag shell over the Au cores), the reaction mixture was cooled to room temperature. The reaction mixture was transferred to the two centrifuge tubes. Then, excess ethanol was added and centrifuged for 5 minutes. The supernatant was poured off, and the precipitate was collected in 4 mL toluene. If isolated Ag nanoparticles were formed, it would show a LSPR absorption peak at around 415 nm. The size-selective precipitation method was used to separate isolated Ag nanoparticles from Au/Ag core/shell NCs.

*Conversion of Au/Ag to Au/$Ag_2S$ Core/Shell NCs*

The procedure reported by Zang et al.[36] was followed to cnvert Au/Ag to Au/$Ag_2S$ core/shell NCs. For a typical conversion, a $1.22*10^{-6}$ M sulfur solution was made by sonicatin under argon. The Au/Ag NCs prepared from the previous step were injected into a one-neck flask. Then, sulfur precursor wad added in the increments of 0.2 mL to Au/Ag NCs solution. In this step, the NCs were vigorously stirred under argon, and left to complete the reactions for ten minutes. After this time, a sample was taken for spectral measurements. In this way, Sulfur was continuously added till the LSPR absorption peak was 630 nm. Further, the NCs were then centrifuged by adding 10 mL ethanol to the reaction mixtures, and the precipitated NCs were dissolved in 4 mL of toluene under argon.

*Conversion of Au/$Ag_2S$ to Au/CdS NCs*

This procedure reported by Zang et al.[36] was taken for this step. In a typical experiment, 0.05g $Cd(NO_3)_2$ was added to 1.0 mL methanol in a one-neck flask, and it was sonicated for 20 minutes. Also, the Au/$Ag_2S$ synthesized from previous step was injected into another flask under argon. After sonication, the $Cd(NO_3)_2$ solution and 0.1 mL TBP were injected in the Au/$Ag_2S$ flask. The reaction temperature was then raised and maintained at 50 C for 45 minutes. It was vigorously stirred during this time. Then, the flask was cooled to room temperature. The reaction mixtures were centrifuged after adding 10 mL methanol, and the precipitate was dissolved in toluene with few drops of oleylamine for colloidal stability.

*Synthesis of CdS NCs*

The CdS NCs were synthesized by using the procedure developed by Yu et al. [37] In three-neck-flask I, 0.0384 g of CdO, 12 mL of ODE, and 0.9 mL of OA were heated under argon at 300 $^{0}$C to dissolve CdO to the solvents. In three-neck-flask II, 0.0098 g sulfur, and 4.5 mL ODE were heated under argon at 200 $^{0}$C to dissolve sulfur and cooled down to 60 $^{0}$C. The flask II mixture was injected to the flask I at 300 $^{0}$C, and cooked for 1.5 minutes, then heating was stopped and cooled down to room temperature. After cooling, the reaction mixture was transferred to centrifuge tubes. An excess of methanol was added and centrifuged for 3 minutes. The clear supernatant was poured off, and CdS NCs were dissolved in hexane. The cleaning with methanol/hexane was repeated again, and the final precipitate was dissolved in hexane.

*Synthesis of Pt NCs*

Platinum NCs were grown using a previously reported protocol.[38] To synthesize Pt NCs, 0.2 mL OA, 0.2 mL OLAM, 43 mg 1,2-hexadecanediol, and 10 mL diphenyl ether were taken in a three-neck-fask. The mixture was degassed at 80 $^{0}$C for one hour. Then, it was switched to argon and the temperature was increased to 200 $^{0}$C. Further, 5mg platinum (II) acetylacetonate was added and heated for 5 minutes till the solution color was turned black. If in 5 minutes, the color of the reaction mixture is not black, additional 5 mg of platinum (II) acetylacetonate was added, and the reaction mixture was heated for 5 more minutes. This step may be repeated as needed and when the black color of the reaction mixture was achieved, the heating was removed from the three-neck-flask and cooled to room temperature. Subsequently, Pt NCs were cleaned two times by precipitating with a methanol.

*Ligands Exchange*

The procedure reported by Costi et al.[39] was adopted to exchange the original hydrophobic ligands with hydrophilic MUA ligands on the NCs. For this, the NCs solution 10-12 mL chloroform was mixed with 10 mg MUA. A KOH solution was prepared by adding 0.1 g KOH in 20 mL ultrapure water. Then, 4 mL KOH solution was added to the NCs mixture and shaken vigourously. Subsequently, the NCs got transferred into the aqueous phase. These NCs were separated and cleaned again using 2 mL of aqueous KOH. Then, the NCs with MUA ligands were precipitated in 10 mL methanol and dissolved in 4 mL distilled water.

*Preparation of Sample Mixtures*

Sample I

| | |
|---|---|
| CdS NCs | 50 unit, |
| AA | 0.08 g, |
| Water | 10 mL, |
| Pt NCs | 0.5 mL, |
| MPA | 0.44 mL, |
| $MV^{2+}$ | 4 drops |

Sample II

| | |
|---|---|
| Au/CdS NCs | 100 unit, |
| AA | 0.08 g |
| Water | 10 mL |
| Pt NCs | 0.5 mL |
| MPA | 0.44 mL |
| $MV^{2+}$ | 4 drops |

*Measurement of Hydrogen Production*

The production of hydrogen with CdS NCs and Au/CdS metal semiconductor NCs were measured under identical conditions. In a typical experiment, two reaction mixtures sample I (50 unit CdS NCs, 0.08 g AA, 10 mL water, 0.5 mL Pt NCs, 0.44 mL MPA, 4 drops $MV^{2+}$) and sample II (100 unit Au/CdS NCs, 0.08 g AA, 10 mL water, 0.5 mL Pt NCs, 0.44 mL MPA, 4 drops $MV^{2+}$) were prepared and the pH was adjusted to 1.5. In a gas tight Shlenk line configuration, a sample containing 10 mL of aqueous solution was freeze-pump-thaw degassed and kept under Ar atmosphere (200 torr). Then, 100 µL headspace volume was taken and injected to the Shimadzu GC-8A with the help of a syringe that gives the base line for the measurement of hydrogen. The sample was then irradiated with the broadband output of a 300 W Xe arc lamp (Oriel) equipped with water filter (to cut IR irradiation) and 400 nm longpass filters for 30 min, and again 100 µL volume was taken and injected to the Shimadzu GC -8A with the help of a syringe. Hydrogen was periodically detected from the headspace using a gas chromatograph equipped with a thermal conductivity detector (Shimadzu GC-8A, argon carrier gas, 5 Å molecular sieve column (Restek), the detector was calibrated against known amounts of $H_2$ gas). In principle, we used excess MPA as the stabilizing ligands and hole scavenger in order to eliminate any role of ligand dissociation and subsequently NCs precipitation on the photocatalytic activity of these materials.

*Characterization*

The absorption and flourscence spectrometry were carried by using CARY 50 scan spectrophotometer and Jobin Yvon Fluorolog FL3-11 fluorescence spectrophotometer respectively. The high resolution TEM images were taken with JEOL 311 UHR that operates at 300 kV. The TEM samples were prepared by depositing a drop of nanoparticle solution in organic solvent onto a carboncoated copper grid and drying it in air. The photocatalytic samples were irradiated with Xe arch lamp (Oriel), and hydrogen produced after irradiation was detected by using Schimadzu GC-8A.

# CHAPTER III: RESULTS AND DISCUSSION

The CdS semiconductor NCs were successfully synthesized as explained in the experimental section. The CdS NCs had an absorption peak at wavelength 430 nm and the transmission electron microscopy (TEM) images (shown in Figure 3.1(b)) are uniform in size. These CdS NCs have lower bandgap than the $TiO_2$ ($E_g$= 3.2 ev) semiconductor. They can absorb visible photons available, and hence, can be used as photon-absorbing materials in photocatalysis of water.

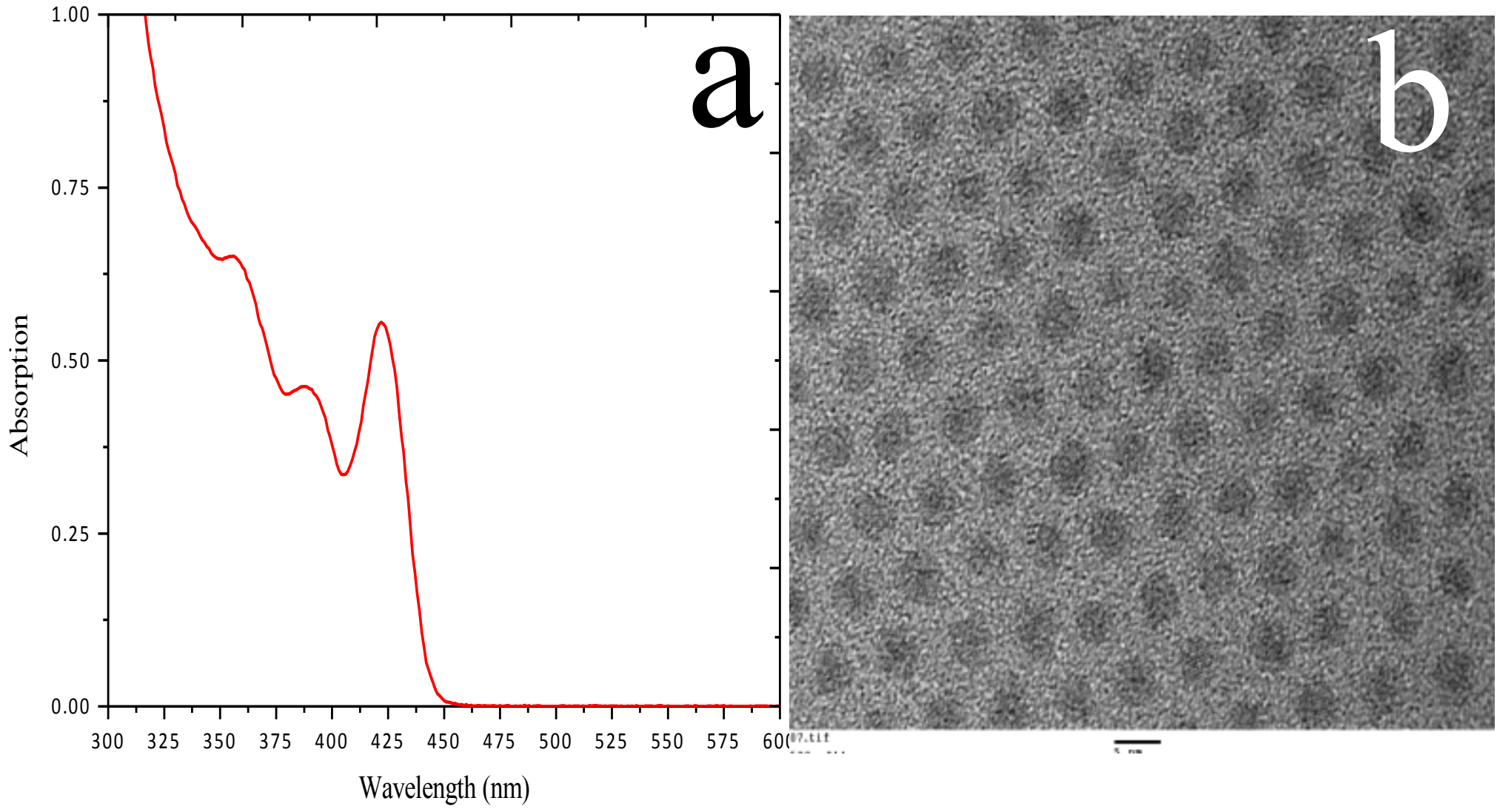

Figure 3.1 Absorption Spectra and TEM Images of CdS NCs

The nonepitaxial Au/CdS core/shell metal semiconductor NCs were synthesized by the four-step synthesis process explained in the experimental section. This method is superior to the previously reported method by Zhang et al.[36] which has multiple steps that takes 24 to 48 hours.

In our study, the synthesis of Au/CdS metal semiconductor NCs followed a simple straightforward process that can be completed in 2 hours with the use of oleylamine (OLAM). The four steps involved in the synthesis of Au/CdS NCs are: 1) synthesis of Au NCs 2) growth of Ag shell on Au NCs 3) conversion of Ag layer into an $Ag_2S$ layer 4) cation exchange to form a CdS shell over the Au NCs. These are schematically represented in Figure 3.2.

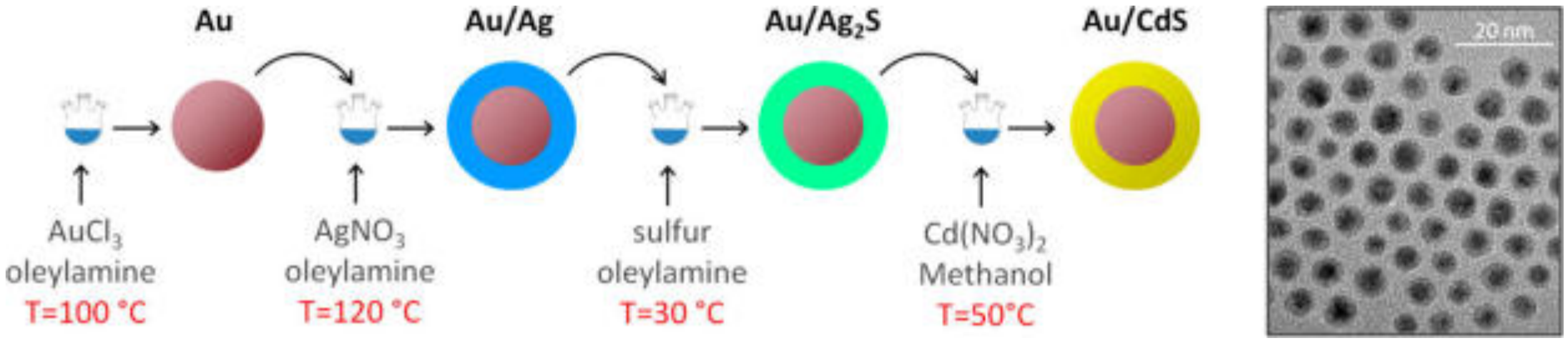

Figure 3.2 A Schematic Representation of the Four Stages in the Synthesis of Au/CdS Metal Semiconductor NCs

The Au NCs were grown differently than the methods reported by the protocols.[36, 40] These Au NCs were synthesized by thermal decomposition of the $AuCl_3$ in OLAM forming primarily monodisperse NCs with a plasmon resonance at around 525 nm. The advantage of this approach is the surface functionalization of Au nanoparticles with OLAM molecules which are suitable surface ligands for the growth of the Ag shell. The Ag layer was deposited by the thermal degradation of $AgNO_3$ on Au surfaces. At this time, OLAM behaves as both the reaction solvent and the capping agent.[36] The $AgNO_3$ solution was added dropwise to avoid the formation of isolated Ag nanoparticles. However, if isolated Ag nanoparticles were formed, then they could be separated from Au/Ag NCs by a size-selective precipitation method. The optical measurements and TEM images showed the formation of Au/Ag core/shell heterostructures.

Figures 3.3(a) and 3.3(f) are the TEM images of the Au NCs with plasmon peak at wavelength 525 nm (Figure 3.3(k)). The addition of $AgNO_3$ solution to the Au NCs formed the Ag layer over the Au NCs and results in blue-shift of the plasmon peak in the heterostructured Au/Ag nanocomposites. The TEM images in Figure 3.3(b) and 3.3(g) confirmed the formation of the Ag layer over the Au NCs. Here, the steady state absorption spectra blueshifted from 525 nm towards the absorption band of Ag nanoparticles (415nm), thus affirming the formation of Ag shell on the Au cores. Sometimes, the sharp absorption peak at 415 nm indicated the formation of isolated Ag nanoparticles during the growth of Au/Ag core/shell heterostructures (shown in Figure 3.3(l)). The isolated Ag nanoparticles could be removed by size-selective precipitation. As reported by the protocol [41], the formation of Au/Ag alloy can be prevented by using low temperature during the synthesis of Au/Ag core/shell. After the growth of the uniform Ag layer, these Ag layers were converted to $Ag_2S$. This process was accompanied by the reaction of Au/Ag core/shell NCs with sulfur at room temperature. At this point, the optical measurements showed that the plasmon resonance shifted toward red from 485 nm to 630 nm. This red-shift is in accordance with Mie theory [42] due to the formation of the dielectric medium over the Au metal cores. In the fourth step, the $Ag_2S$ layer was converted to CdS by the cation exchange method. Here, the plasmon peak blueshifted from 630 nm to 590 nm. In addition, a weak absorbance at around 460 nm corresponds to ground state transitions of CdS semiconductor domains. The change in the dielectric constant surrounding the Au cores is responsible for the blue-shift of the plasmon peak. The TEM images of these Au/CdS cores/shells indicates that the CdS layer is uniform over the Au cores. Images in Figures 3.3(d) and 3.3(j) correspond to 16.1 nm core

domains (CdS thickness≈4.1nm) and Figures 3.3(e) and 3.3(i) correspond to a 6.5 nm core domain (CdS thickness ≈ 3.9 nm). The CdS shells over the Au cores were adequately uniform.

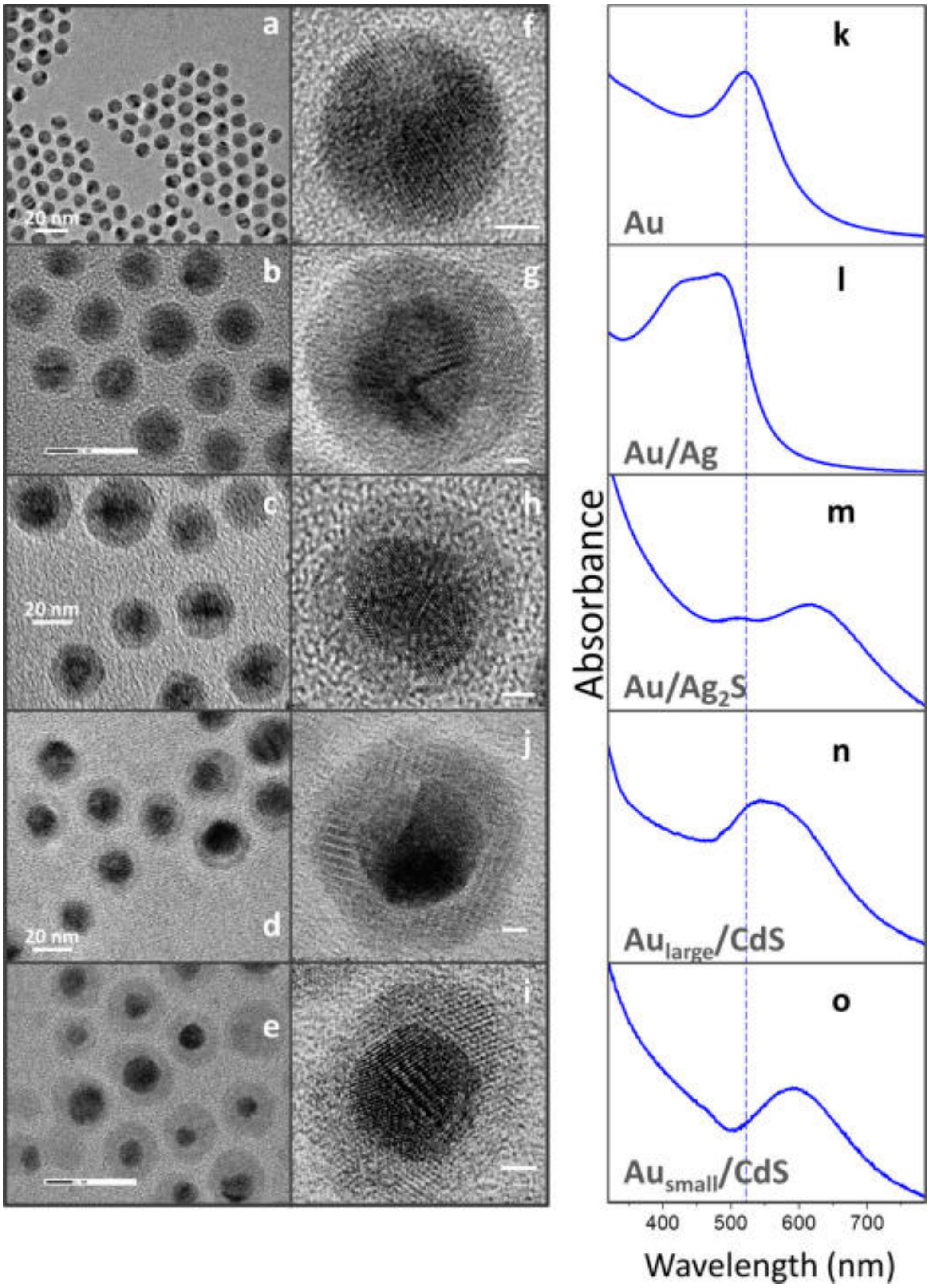

Figure 3.3 TEM Images and Absorption Spectra of Nanocomposites Formed during each of the Four Reaction Steps Involved in the Synthesis of the Au/CdS NCs

The CdS semiconductor NCs and Au/CdS metal semiconductor NCs have different light absorption capacities. The Au/CdS NCs absorb more light than the CdS NCs. This can be demonstrated by a simple test where two different NCs with the same absorption coeffcient can be deposited to TEM grid to compare the ratio of quantities of one type of NCs versus the other. The Figure 3.4(a) represents the absorption spectra of CdS NCs and Au/CdS NCs and Figures 3.4(b) and 3.4(c) are the TEM images of the respective NCs. (See Figure 3.4 below). Although these two different NCs have same absorption extinction coefficients, the number of particles in CdS NCs is significantly higher than the number of Au/CdS NCs, which indicates that more number of CdS NCs are needed to absorb the same amount of light as Au/CdS metal semiconductor NCs. The ratio of number of CdS NCs to Au/CdS NCs is around 20:1. It implies that there should be 20 times greater number of CdS NCs than the Au/CdS NCs to absorb same amount of light. Further, the higher light absorption by metal semiconductor NCs than semiconductor NCs is due to the larger absorption cross-section of the metal semiconductor NCs. Thus, the presence of Au metal in the semiconductor NCs significantly enhances the light absorption due to the presence of plasmon resonance.

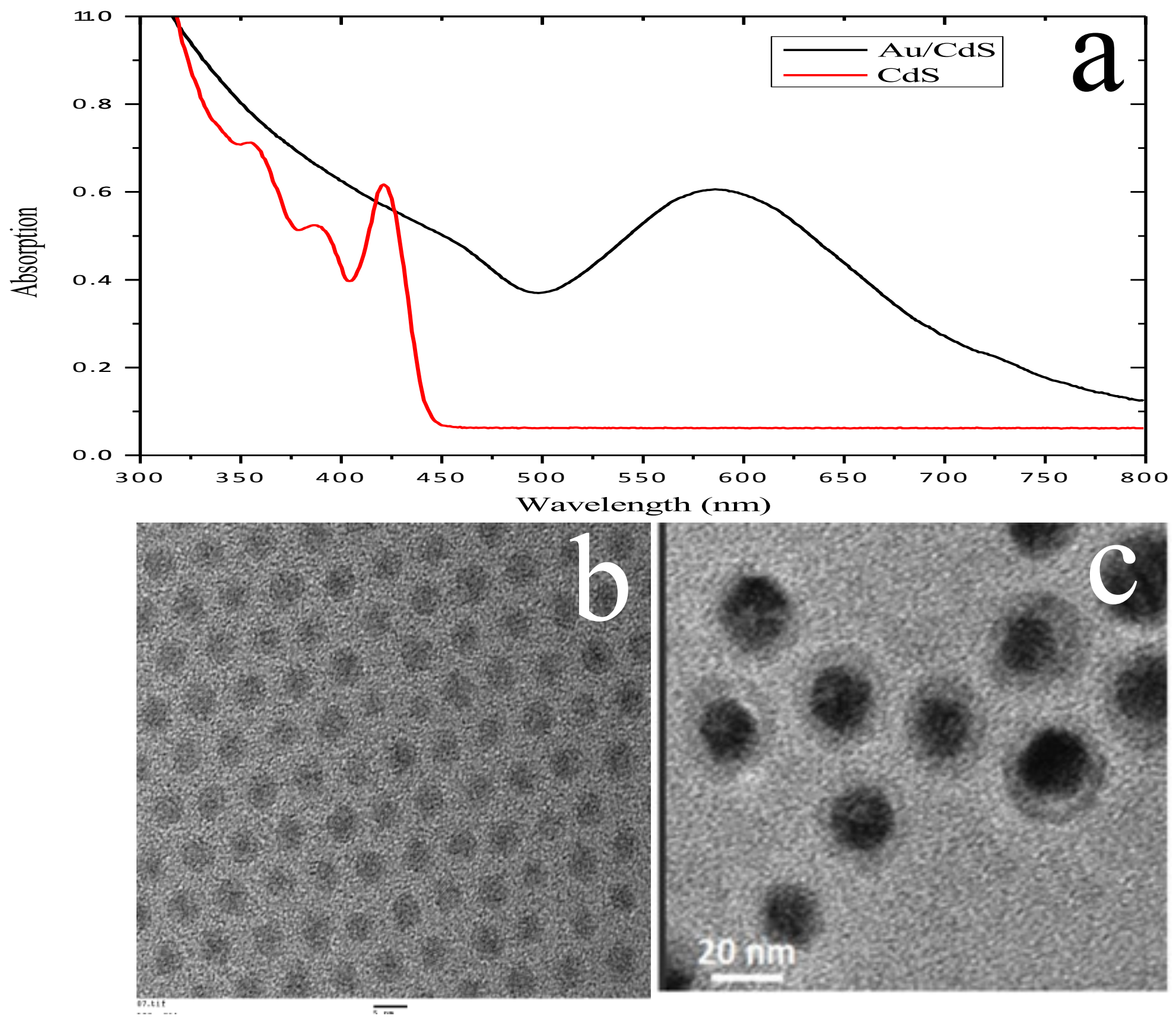

Figure 3.4 Absorption Spectra and TEM Images of CdS and Au/CdS NCs

To make a homogeneous system for photoccatalysis of water, OA and OLAM ligands on CdS ans Au/CdS respectively were exchanged to MUA ligands to make them water-soluble. Then, the photochemical experiments were conducted by the illumination of standard solar light. In the photocatalysis of water, two water molecules split into two $H_2$ molecules (reduction part) and one $O_2$ molecule (oxidation part). Electrons from the oxidative part are routed to the reductive part where they reduce protons to hydrogen. Though it is a complex to analyze the reduction and oxidation reactions, the study of the reduction half reaction of protons is common for a sacrificial electron donor to be used to optimize catalysts for hydrogen production.[1] Since the reduction of protons by ascorbate [E = -0.41 V] is not favorable, ascorbic acid (AA) was used as a sacrificial electron donor in the homogeneous system.

In a typical experiment, the homogeneous system was irradiated with the standard solar light and hydrogen was produced by a solution of NCs, AA, water, Pt NCs (or $NiNO_3$ solution), MPA, and $MV^{2+}$. Here, in this homogeneous system, AA donates sacrificial electrons and behaves as the buffer solution to maintain the concentration of $H^+$ ion in the system. NCs absorb the photons, it creates electron-hole pairs. The excited electrons transfer to the Pt NCs, and they reduce protons, and hence, produce hydrogen. The MPA ligands scavenge holes, thus suppressing electron-hole recombinations.

The photocatalytic activity of CdS (Au/CdS) NCs can be described with the help of the schematic diagram shown below. As the photons incident on the NCs surface, then the catalytic activity follows a general strategy that have a) absorption of visible photons to produce excited electron-hole pairs, b) migration of electrons to catalysts where they are used for reduction reactions, and c) scavenging of holes to refill the electrons utilized for reduction and to prevent NCs degradation due to oxidizing holes.[3] Here, after the photoexcitation of electrosn, these electrons may combine with the hole on the valence band and release the photonic energy. The direct measurements of charge-transfer dynamics in colloidal CdS-Pt nano-heterostructures showed that energy transfer from CdS to Pt catalyst happened faster than electron-hole recombinations.[43- 45] Energy transfer from CdS to Pt was fast (≈ 3 ps) in CdS-Pt interface and the charge separated-state was long-lived (≈ 1 μs) due to hole trapping at the CdS surface [12]. Not only NCs properties but also the size of Pt catalysts play an important role in photocatalysis of water. [46] The study shows that the small Pt dots are better suited for the photocatalytic reduction of protons. [47] Further, the ligands on the NCs surface scavenge holes, and the addition of ligand molecules (MUA or MPA) enhances the $H_2$ generation rates by suppressing the electron-hole recombination processes.

We hypothesize that the catalytic activity begins after the light absorption by the CdS (Au/CdS) NCs; then the electrons are excited. These electrons transfer to the surface of the catalysts, transferring the photonic energy, available in the solution and hence, there the protons are reduced. The oxidation of ascorbic acid (AA) produces holes, $2e^-$ , and $2H^+$ to the homogeneous photocatalytic system. Here, AA thus behaves as both a potential source of hydrogen and as a buffer since it creates a conjugate base and maintains the pH though the protons are reduced to hydrogen. As the concentration of $H^+$ decreases in the system, the water molecules dissociate and hence its concentration is maintained.

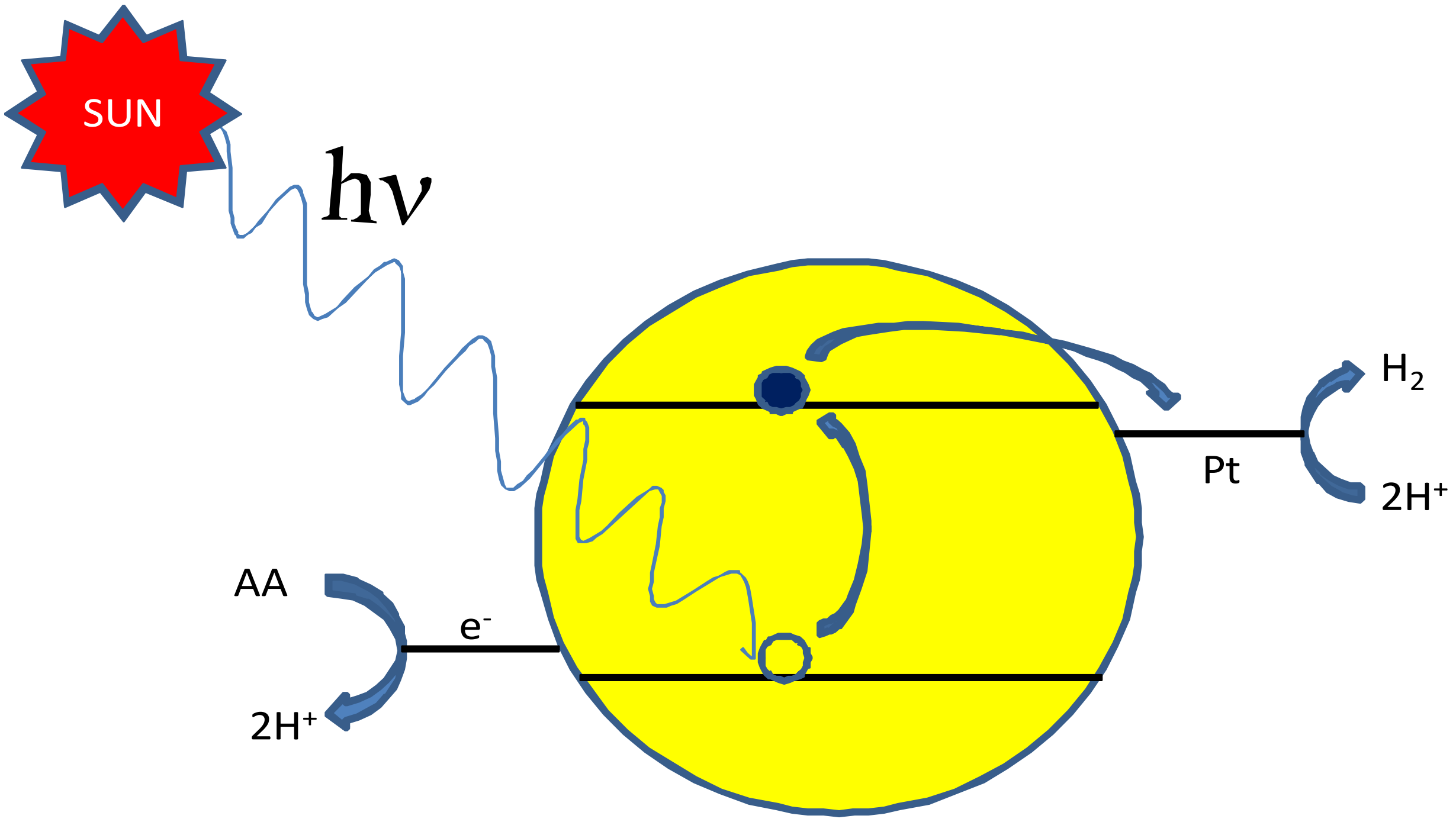

Figure 3.5 Schematic Representation of Photocatalysis Using Semiconductor NCs

Initially, to study the photocatalytic activity of Au/CdS metal semiconductor NCs, the sample was prepared as stated in the experimental section. Here, $NiNO_3$ was used as the catalysts that accept the excited electrons to reduce protons in the homogeneous solution. It was irradiated for 30 minutes, and the hydrogen produced was detected with Shimadzu GC-8A detector. The graph (Figure 3.6) below shows the amount of hydrogen produced. Though oxygen is produced during the catalysis of water, only the hydrogen produced was detected. Hence, we conclude that Au/CdS NCs, they absorb light significantly, and hence, they reduce hydrogen via the energy transfer.

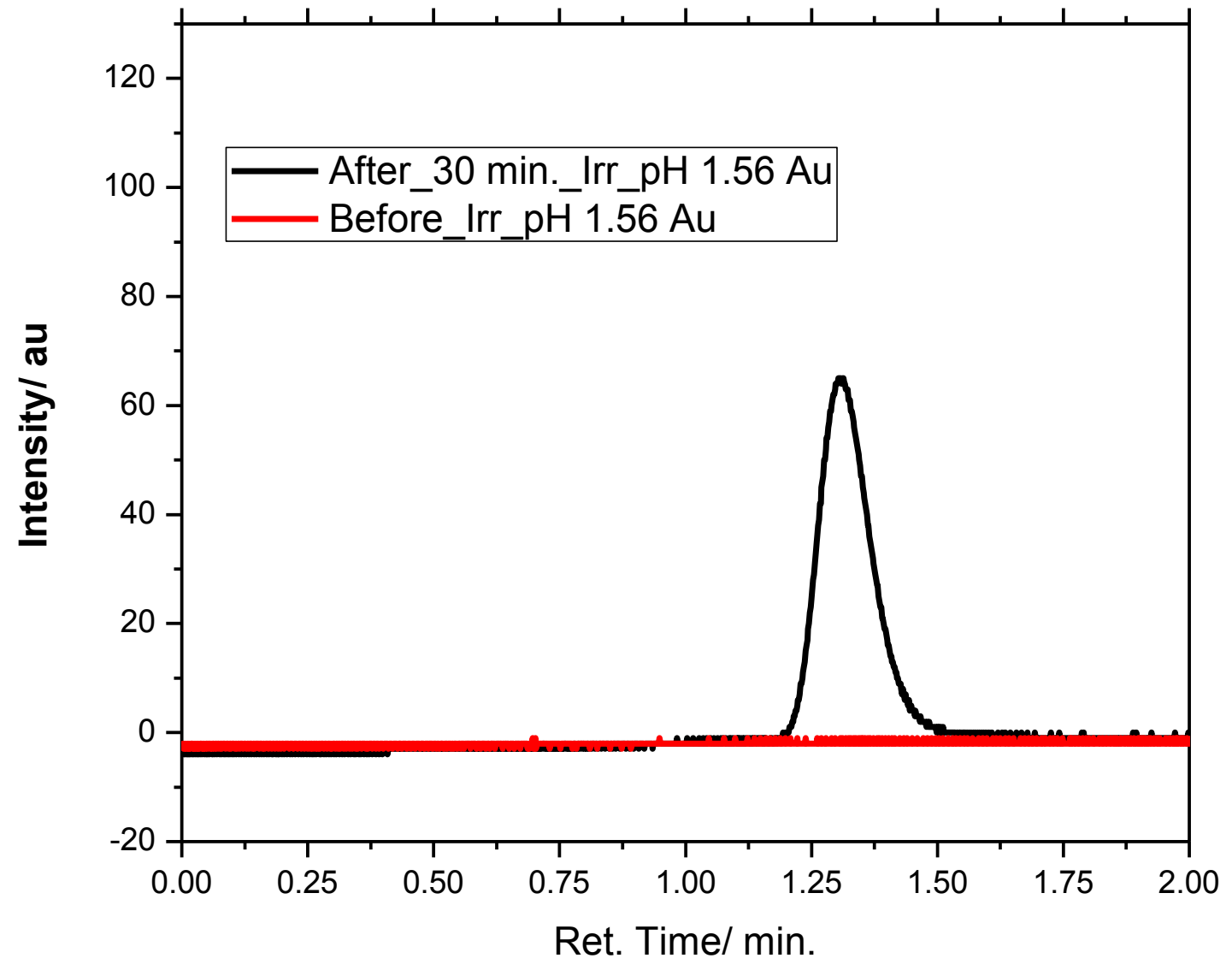

Figure 3.6 Graphical Representation of Hydrogen Produced by Au/CdS NCs with Nickel Nitrate ($NiNO_3$) Catalysts

Similarly, we compared the catalytic activity of Au/CdS metal semiconductor NCs with CdS semiconductor NCs under identical conditions. At this time, Pt NCs were used as the catalysts rather than $NiNO_3$ since the CdS NCs form CdS-Pt cluster that helps to transfer the electrons from conduction band to the surface of catalysts.[12] Thus, it would increase rate of hydrogen reduction. Now, to investigate the photocatalytic activity of CdS semiconductor NCs and Au/CdS metal semiconductor NCs, the samples were prepared as discussed in the methods section. These two samples of different NCs were irradiated with the standard solar light. They were kept under illumination for same time in identical conditions, and the amounts of hydrogen produced were measured. Figures 3.7 and 3.8 respectively represent the hydrogen produced in the typical experiment using CdS and Au/CdS NCs. Figure 3.9 represents the comparison between the amounts of hydrogen produced by CdS and Au/CdS NCs respectively.

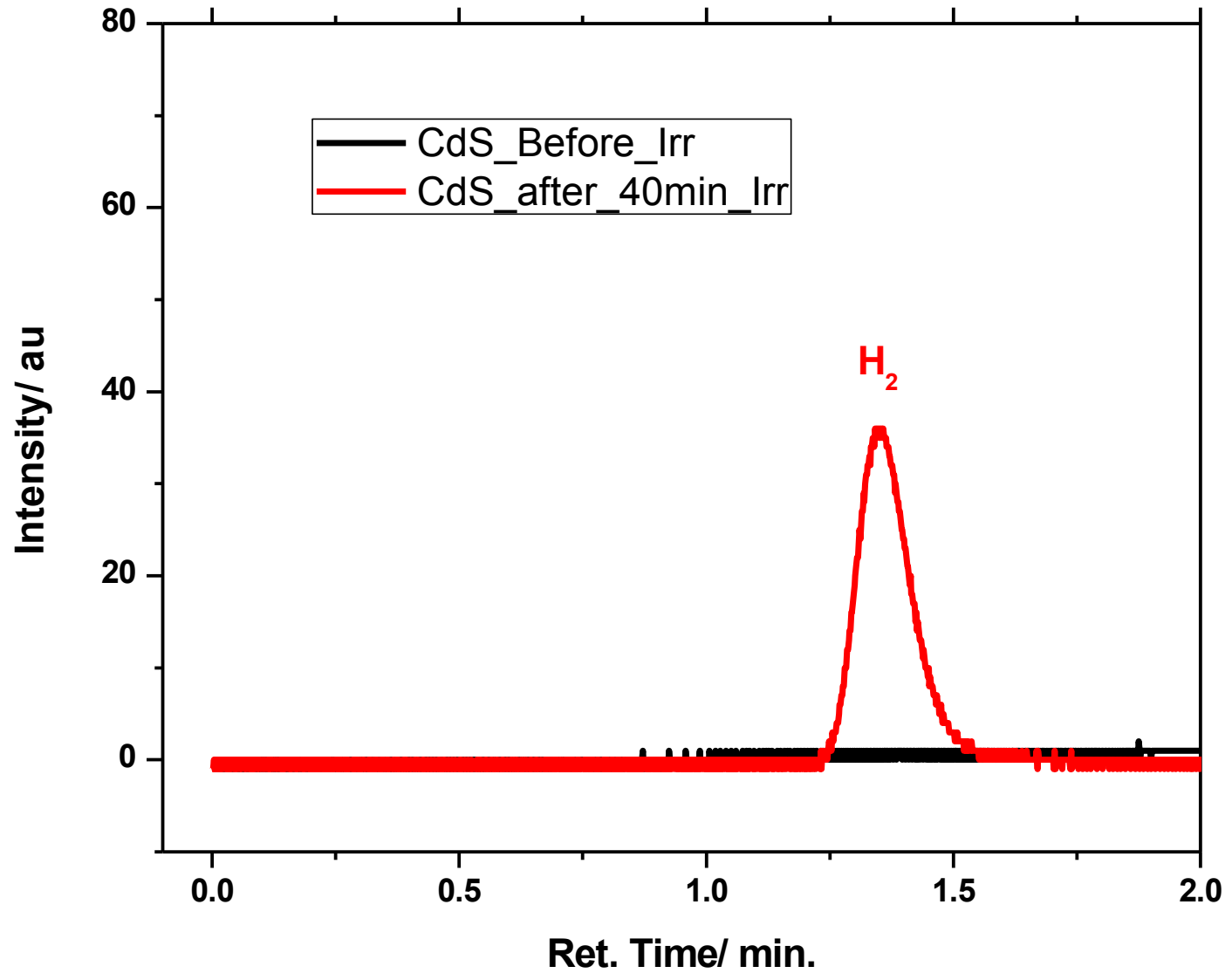

Figure 3.7 Graphical Representation of Hydrogen Produced by CdS NCs

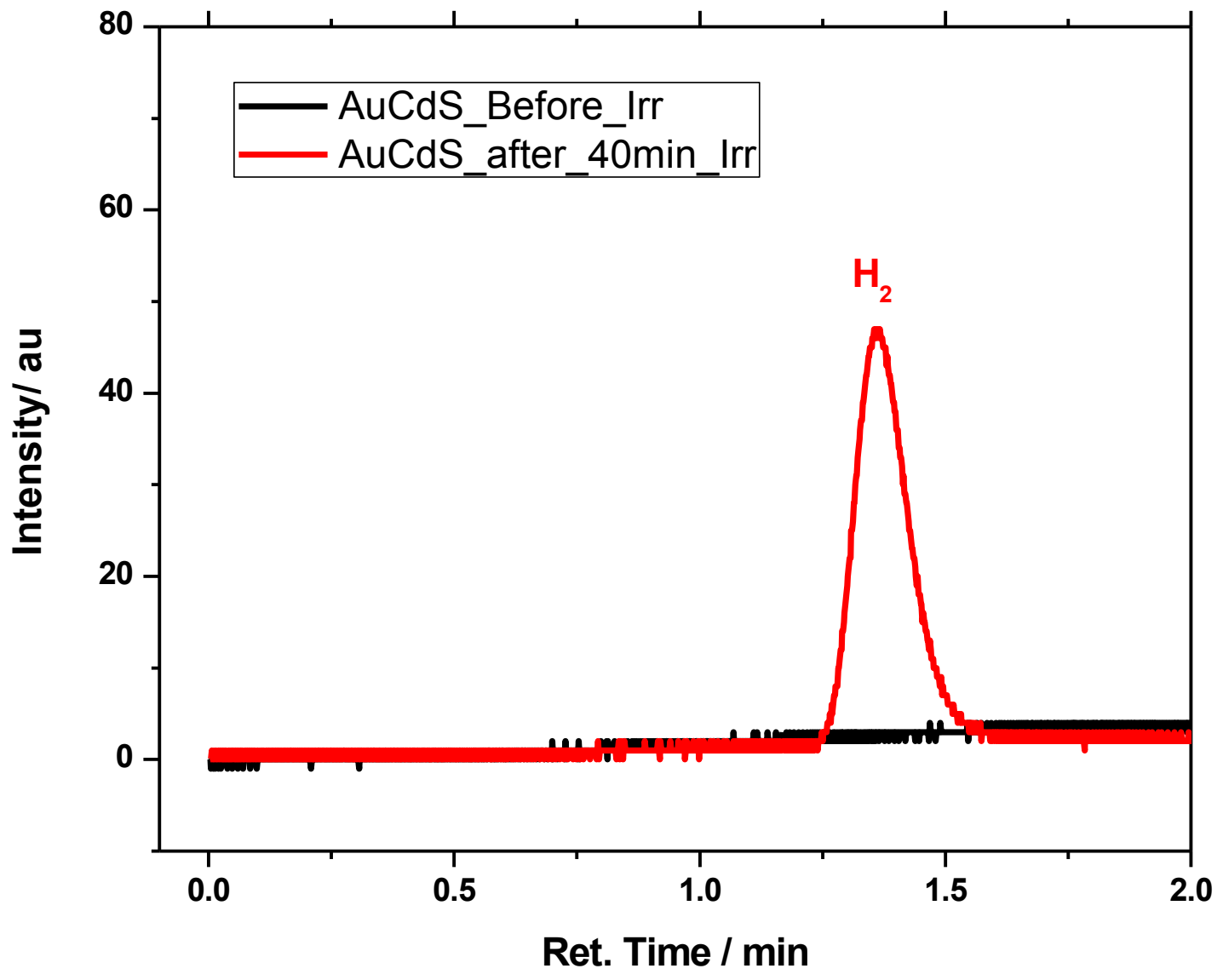

Figure 3.8 Graphical Representation of Hydrogen Produced by Au/CdS NCs

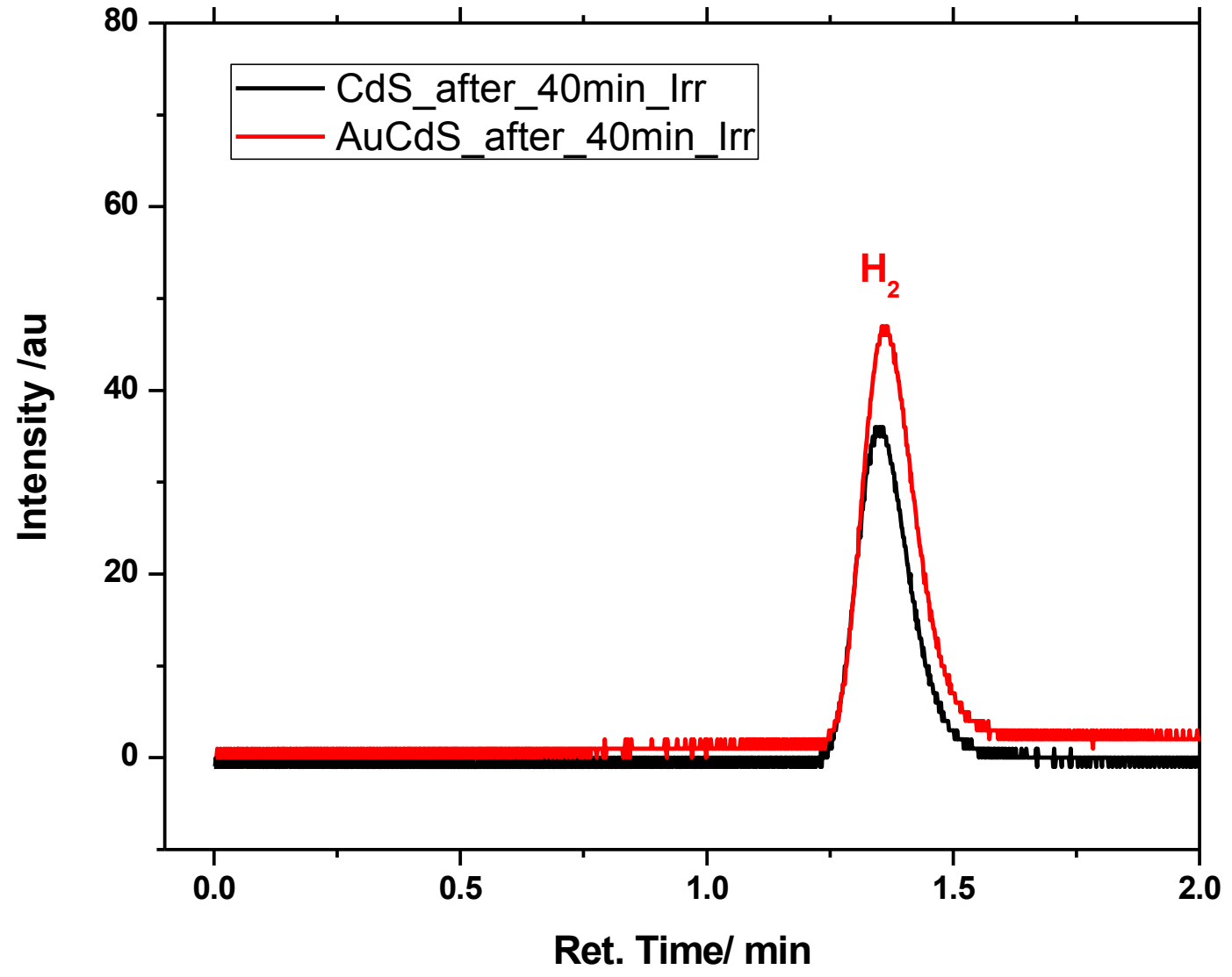

Figure 3.9 Comparison of Amounts of Hydrogen Produced by CdS and Au/CdS NCs

Here, the amount of hydrogen produced by Au/CdS is greater than the CdS NCs though the number of Au/CdS NCs used was less than the number of CdS NCs. It is due to the absorption of large amount of photons by the Au/CdS NCs. Though all the absorbed photons do not migrate to the catalysts available, it produces large amount of hydrogen as compared to the CdS NCs. It requires further investigation to compare the amounts of hydrogen produced by Au/CdS and CdS NCs quantitatively. Thus, it confirms that the photons absorption by metal semiconductor NCs is greatly enhanced by plasmon resonance, and hence, they can enhance the catalytic activity of semiconductor NCs by energy transfer from metal into semiconductor material.

# CHAPTER IV: CONCLUSION

In summary, the Au/CdS NCs were successfully synthesized, and the catalytic activity of Au/CdS metal semiconductor NCs was studied. These Au/CdS metal semiconductor NCs NCs absorb significantly larger amount of visible photons as compared to CdS NCs due to the plasmon-resoance. Hence, these Au/CdS metal semiconductor transfer energy from metal to semiconductor materials that finally reduces protons into hydrogen. This project work reveals that the hydrogen production by using metal semiconductor NCs (Au/CdS) is more efficient than that of using semiconductor NCs (CdS) under identical conditions, and it requires further investigation for quantitative analysis. Further, the addition of other ligands, such as MPA, helps to scavenge holes, and hence suppresses electro-hole recombination processes.

# REFERENCES

1. Han, Z., Qiu, F., Eisenberg, R., Holland, P. L., & Krauss, T. D. (2012). Robust photogeneration of $H_2$ in water using semiconductor nanocrystals and a nickel catalyst. *Science, 338*(6112), 1321-1324.

2. Zhu, H., Song, N., Lv, H., Hill, C., & Lian, T. (2012). Near unity quantum yield of light-driven redox mediator reduction and efficient $H_2$ generation using colloidal nanorod heterostructures. *Journal of the American Chemical Society, 134*(28), 11701-11708.

3. Wilker, M. B., Schnitzenbaumer, K. J., & Dukovic, G. (2012). Recent progress in photocatalysis mediated by colloidal II-VI nanocrystals. *Israel Journal of Chemistry, 52*(11), 1002-1015.

4. Acharya, K. P., Zamkov, M., Khnayzer, R. S., O'Connor, T., Diederich, G., Kirsanova, M., et al. (2011). The role of hole localization in sacrificial hydrogen production by semiconductor-metal heterostructured nanocrystals. *Nano Letters, 11*(7), 2919-2926.

5. Hoffmann, M., Martin, S., Choi, W., & Bahnemann, D. (1995). Environmental applications of semiconductor photocatalysis. *Chemical Reviews, 95*(1), 69-96.

6. Mills, A., & Le Hunte, S. (1997). An overview of semiconductor photocatalysis. *Journal of Photochemistry & Photobiology, A: Chemistry, 108*(1), 1-35.

7. Rosenthal, S. J., McBride, J., Pennycook, S. J., & Feldman, L. C. (2007). Synthesis, surface studies, composition and structural characterization of CdSe, core/shell and biologically active nanocrystals. *Surface Science Reports, 62*(4), 111-157.

8. Parak, W. J., Manna, L., Simmel, F. C., Gerion, D., & Alivisatos, P. (2003; 2005). Quantum dots. *Nanoparticles* (pp. 4-49) Wiley-VCH Verlag GmbH & Co. KGaA.

9. Rogach, A. L. (2008). *Semiconductor nanocrystal quantum dots: Synthesis, assembly, spectroscopy and applications*. Wien: Springer.

10. Zeng, S., Yong, K., Roy, I., Dinh, X., Yu, X., & Luan, F. (2011). A review on functionalized gold nanoparticles for biosensing applications. *Plasmonics, 6*(3), 491-506.

11. Lambright, S., Butaeva, E., Razgoniaeva, N., Hopkins, T., Smith, B., Perera, D., et al. (2014). Enhanced lifetime of excitons in nonepitaxial Au/CdS Core/Shell nanocrystals. *ACS Nano, 8*(1), 352-361.

12. Willets, K. A., & Van Duyne, R. P. (2007). Localized surface plasmon resonance spectroscopy and sensing. *Annual Review of Physical Chemistry, 58*(1), 267-297.

13. Reiss, P., Protière, M., & Li, L. (2009). Core/Shell semiconductor nanocrystals. *Small, 5*(2), 154-168.

14. Xie, R., Zhong, X., & Basché, T. (2005). Synthesis, characterization, and spectroscopy of type-II Core/Shell semiconductor nanocrystals with ZnTe cores. *Advanced Materials, 17*(22), 2741-2745.

15. Dickerson, B. D. (2005). *Organometallic Synthesis Kinetics of CdSe Quantum Dots.*

16. Smith, A., Dave, S., Nie, S., True, L., & Gao, X. (2006). Multicolor quantum dots for molecular diagnostics of cancer. *Expert Review of Molecular Diagnostics, 6*(2), 231-244.

17. Kairdolf, B. A., Smith, A. M., Stokes, T. H., Wang, M. D., Young, A. N., & Nie, S. (2013). Semiconductor quantum dots for bioimaging and biodiagnostic applications. *Annual Review of Analytical Chemistry, 6*(1), 143-162.

18. Henry, N. L., & Hayes, D. F. (2006). Uses and abuses of tumor markers in the diagnosis, monitoring, and treatment of primary and metastatic breast cancer. *The Oncologist, 11*(6), 541-552.

19. Voura, E. B., Jaiswal, J. K., Mattoussi, H., & Simon, S. M. (2004). Tracking metastatic tumor cell extravasation with quantum dot nanocrystals and fluorescence emission-scanning microscopy. *Nature Medicine, 10*(9), 993-998.

20. Rogach, A. L. (2008). *Semiconductor nanocrystal quantum dots: Synthesis, assembly, spectroscopy and applications* Springer Verlag, Wien.

21. Sargent, E. (2012). Colloidal quantum dot solar cells. *Nature Photonics, 6*(3), 133-135.

22. Kinder, E., Moroz, P., Diederich, G., Johnson, A., Kirsanova, M., Nemchinov, A., et al. (2011). Fabrication of all-inorganic nanocrystal solids through matrix encapsulation of nanocrystal arrays. *Journal of the American Chemical Society, 133*(50), 20488-20499.

23. https://life.llnl.gov/why_life/index.php

24. Daneshwar, N., Aber, S., Seyed Dorraji, M. S., Khataee, A. R., & Rasoulifard , M. H. (2007). Preparation and Investigation of Photocatalytic Properties of ZnO Nanocrystals: Effect of Operational Parameters and Kinetic Study. *International Journal of Chemical, Materials Science and Engineering,* Vol. 1, No. 5, 71-76.

25. Fujishima, A., & Honda, K. (1972). Electrochemical Photolysis of Water at a Semiconductor Electrode. *Nature, 238*(5358), 37-37.

26. Chen, X., Liu, L., Yu, P. Y., & Mao, S. S. (2011). Increasing solar absorption for photocatalysis with black hydrogenated titanium dioxide nanocrystals. *Science (New York, N.Y.), 331*(6018), 746-750.

27. Waterhouse, G., Wahab, A., Al-Oufi, M., Jovic, V., Anjum, D., Sun-Waterhouse, D., et al. (2013). Hydrogen production by tuning the photonic band gap with the electronic band gap of TiO2. *Scientific Reports, 3*.

28. Borse, P.H., Janj, J. S., Hong, S. J., Lee, J. S., et al. (2009). Photocatalytic Hydrogen Generation from Water-methanol Mixtures Using Nanocrystalline $ZnFe_2O_4$ under Visible Light Irradiation. *Journal of the Korean Physical Society*, Vol. 55, No. 4, October 2009 pp. 1472-1477.

29. Sun, G., Zhang, S., & Li, Y. (2014). Solvothermal Synthesis of $Zn_2SnO_4$ Nanocrystals and Their Photocatalytic Properties. *International Journal of Photoenergy*, vol. 2014, Article ID 580615, 1-7.

30. Liao, L., Yu, Q., Cai, X., Zhao, J., Ren, Z., Fang, H., et al. (2014). Efficient solar water-splitting using a nanocrystalline CoO photocatalyst. *Nature Nanotechnology, 9*(1), 69-73.

31. O' Connor, T., Panov, M. S., Mereshchenko, A., Tarnovsky, A. N., Lorek, R., Perera, D., et al. (2012). The effect of the charge-separating interface on exciton dynamics in photocatalytic colloidal heteronanocrystals. *ACS Nano, 6*(9), 8156-8165.

32. Khon, E., Lambright, K., Khnayzer, R. S., Moroz, P., Perera, D., Butaeva, E., et al. (2013). Improving the catalytic activity of semiconductor nanocrystals through selective domain etching. *Nano Letters, 13*(5), 2016-2023.

33. Zhao, J., Holmes, M. A., & Osterloh, F. E. (2013). Quantum confinement controls photocatalysis: A free energy analysis for photocatalytic proton reduction at CdSe nanocrystals. *ACS Nano, 7*(5), 4316-4325.

34. Khon, E., Hewa-Kasakarage, N., Nemitz, I., Acharya, K., & Zamkov, M. (2010). Tuning the morphology of Au/CdS nanocomposites through temperature-controlled reduction of gold-oleate complexes. *Chemistry of Materials, 22*(21), 5929-5936.

35. Shore, M. S., Wang, J., Johnston-Peck, A. C., Oldenburg, A. L., & Tracy, J. B. (2011). Synthesis of Au(core)/Ag(shell) nanoparticles and their conversion to AuAg alloy nanoparticles. *Small (Weinheim an Der Bergstrasse, Germany), 7*(2), 230-234.

36. Zhang, J., Tang, Y., Lee, K., & Ouyang, M. (2010). Nonepitaxial growth of hybrid core-shell nanostructures with large lattice mismatches. *Science, 327*(5973), 1634-1638.

37. Yu, W. W., & Peng, X. (2002). Formation of high-quality CdS and other II-VI semiconductor nanocrystals in noncoordinating solvents: Tunable reactivity of monomers. *Angewandte Chemie (International Ed.in English),41*(13), 2368-2371.

38. Habas, S. E., Yang, P., & Mokari, T. (2008). Selective growth of metal and binary metal tips on CdS nanorods. *Journal of the American Chemical Society, 130*(11), 3294-3294.

39. Costi, R., Saunders, A. E., Elmalem, E., Salant, A., & Banin, U. (2008). Visible light-induced charge retention and photocatalysis with hybrid CdSe-Au nanodumbbells. *Nano Letters, 8*(2), 637-641.

40. Zhang, J., Tang, Y., Lee, K., & Ouyang, M. (2010). Tailoring light-matter-spin interactions in colloidal heteronanostructures. *Nature, 466*(7302), 91-95.

41. Sagarzazu, G., Inoue, K., Saruyama, M., Sakamoto, M., et al. (2013). Ultrafast Dynamics and Single Particle Spectroscopy of Au-CdSe Nanorods. *Phys. Chem. Chem. Phys*. 15, 2141–2152.

42. Mie, G. (1908). Contributions to the Optics of Diffuse Media, Especially Colloid Metal Solutions. *Ann Phys*. 25, 377–445.

43. Berr, M. J., Vaneski, A., Mauser, C., & Fischbach, S. (2012). Delayed photoelectron transfer in pt-decorated CdS nanorods under hydrogen generation conditions. *Small (Weinheim an Der Bergstrasse, Germany), 8*(2), 291-297.

44. Wu, K., Zhu, H., Liu, Z., Rodriguez-Cordoba, W., & Lian, T. (2012). Ultrafast charge separation and long-lived charge separated state in photocatalytic CdS-pt nanorod heterostructures. *Journal of the American Chemical Society, 134*(25), 10337-10340.

45. Osterloh, F., & Parkinson, B. (2011). Recent developments in solar water-splitting photocatalysis. *MRS Bulletin, 36*(1), 17-22.

46. Berr, M., Vaneski, A., Susha, A. S., et al. (2010). Colloidal CdS nanorods decorated with subnanometer sized Pt clusters for photocatalytic hydrogen generation. *Applied Physics Letters, 97*(9), 093108, 1-3.

47. Elmalem, E., Saunders, A. E., Costi, R., & Salant, A. (2008). Growth of photocatalytic CdSe-pt nanorods and nanonets. *Advanced Materials (Weinheim), 20*(22), 4312- 4317.